\documentstyle[12pt,epsfig]{article}
\begin{document}

\begin{titlepage}
\rightline{April 2004}
\vskip 3cm
\centerline{\Large \bf
Testing the mirror world hypothesis
}
\vskip 0.5cm
\centerline{\Large \bf
for the close-in extrasolar planets}

\vskip 2.2cm
\centerline{R. Foot\footnote{
E-mail address: foot@physics.unimelb.edu.au}}

\vskip 0.7cm
\centerline{\it School of Physics}
\centerline{\it University of Melbourne}
\centerline{\it Victoria 3010 Australia}
\vskip 2cm
\noindent
Because planets are not expected to be able to 
form close to stars due to the high temperatures,
it has been suggested that the observed close orbiting ($\sim 0.05$
AU) large mass planets ($\sim M_J$)
might be mirror worlds -- planets composed predominately of mirror matter.
The accretion of ordinary matter onto the mirror planet (from e.g.
the solar wind from the host star) will make the mirror planet
opaque to ordinary radiation with an
effective radius $R_p$. It was argued in a previous paper,
astro-ph/0101055, that this radius was
potentially large enough to explain the 
measured size of the first transiting close-in extrasolar planet, HD209458b.
Furthermore, astro-ph/0101055,
made the rough prediction: $R_p \propto \sqrt{{T_s \over M_p}}$,
where $T_s$ is the surface temperature of the ordinary matter in the
mirror planet and $M_p$ is the mass of the planet (the
latter dependence being the more robust prediction). We compare
this prediction with the recently discovered transiting
planets, OGLE-TR-56b and OGLE-TR-113b.

\end{titlepage}
Since the 1995 discovery of a planet around the 
star 51 Pegasi\cite{swiss},
more than 100 extrasolar planets have been found\cite{enc}.
Perhaps the most surprising characteristic of these 
planets is that some of them (including 15 Pegasi b)
have been found which
have large mass ($\sim M_J$)
with orbits very close to their star ($\sim 0.05$ AU). 
This is surprising because the environment close to the
star is far too hot for giant planet formation to
occur\cite{boss}.

Some years ago, it was suggested\cite{f1,f2} that the close-in
extrasolar planets might be composed not of ordinary
matter, but primarily of mirror matter\cite{mm}. 
If this were 
the case, then the close-in orbits of the planets would
not pose any problem.
A significant ordinary matter subcomponent ($\sim 10^{-3} M_J$) would
necessarily occur -- being accreted from the stellar
wind of the host star\cite{f2}. 
The ordinary matter subcomponent will make the mirror
planet opaque to ordinary radiation with an
effective radius, $R_p$.
The ordinary matter
subcomponent would be very hot and (relatively) low
in density. This means that the ideal gas law 
could be used to relate the pressure to the density and
temperature
and from the condition of hydrostatic
equilibrium a simple relation could be derived for
$R_p$\cite{f2}:
\begin{eqnarray}
R_p \propto \sqrt{{T_s \over M_p} }
\label{1}
\end{eqnarray}
where $T_s$ is the surface temperature of the planet and
$M_p$ is the mass of the planet.
This was only a rough prediction (especially the
dependence on $T_s$) but a prediction nevertheless.
Heuristically it is very easy to understand: increasing the mass
$M_p$ increases the force of gravity which causes the gas of
ordinary matter to become more tightly bound to the
mirror planet (thereby decreasing the effective size, $R_p$),
while increasing the temperature of the gas increases the volume
that the gas occupies (thereby increasing $R_p$).
Of these two effects we expect that the dependence on $M_p$ should
be the more robust prediction. 
Because the size of ordinary gas giant planets (i.e. planets
made mostly of ordinary matter) depends quite weakly on their
mass the 
dependence on $M_p$, which is significant according to Eq.(\ref{1}), 
should allow a 
decisive test of
the mirror planet hypothesis. 

Although
radial velocity surveys have been very successful in
finding planetary candidates, they do not give the
mass of the planet, only $M_p \sin i$, and do not
provide information about the planet's size, $R_p$.
This information can be obtained if transiting
systems are found. In this case $\sin i \simeq 1$ (which
means that the mass of the planet can be established
from the radial velocity measurements), and the
depth of the transit 
can be used to estimate the planets size.
The first transiting planet, HD209458b was discovered in
1999\cite{henry,char,maz}, with
parameters
$M_p = 0.69 \pm 0.05 M_J$ and $R_p = 1.43\pm 0.05 R_J$\cite{deeg}.
The large size of this planet indicates that the interior
is very hot, with some authors suggesting that
some internal heating mechanism is required\cite{bod}
while others argue that this is not necessary\cite{bur}.

Following the discovery of the first transiting planet,
HD209458b, intensive efforts have been underway
to search for other transiting extrasolar planets.
The OGLE survey (Optical Gravitational Lensing Experiment) 
announced the detection of more than 100 short-period
transiting objects\cite{ogle}.
These observations were followed up with radial velocity 
measurments leading to the recent discovery of three new 
transiting planets:
OGLE-TR-56b\cite{kon}, OGLE-TR-113b\cite{bou,kon4} and OGLE-TR-132b\cite{bou}.
It will be interesting to see how well the rough prediction, Eq.(\ref{1}),
agrees with this new data.

The effective surface temperature of the planet, $T_s$, can be related to that
of the star via:
\begin{eqnarray}
T_s = \left( {1-a\over 4}\right)^{1/4} 
\left( {R_{star} \over D}\right)^{1/2}
\ T_{eff}^{star}
\label{2}
\end{eqnarray}
where $a$ is the albedo, $R_{star}$ is the radius of the star
and $D$ is the (mean) orbital distance of the planet.
The parameters $D, \ R_{star}, \ T_{eff}^{star}$ for the systems
with transiting planets are given in table 1.
\begin{table}[h]
\vspace{3mm}
{\centering
\begin{tabular}[t]{|c|c|c|c|}
\hline
{\textbf{Transiting planet}} &
${D} \ [AU] $ & $R_{star} \ [R_{\odot}]$ & $T_{eff}^{star}$\\
\hline
\hline
{HD209458b} &
0.045\cite{henry}
& {\raggedright $1.20\pm 0.02$\cite{deeg}}
& {\raggedright $6000 \pm 50$\ K\cite{maz}} \\
\hline
{OGLE-TR-56b} &
{\raggedright $0.0225$\cite{kon} }& {\raggedright $1.10 \pm 0.10$\cite{sas}}
& $5900 \pm 80$\ K \cite{sas}\\
\hline
{OGLE-TR-113b} &
{\raggedright $0.0228$\cite{bou}}& {\raggedright
$0.765 \pm 0.025$\cite{bou}}&{\raggedright $4752 \pm 130 \ K$\cite{bou}}
\\
\hline
{OGLE-TR-132b} &
{\raggedright $0.0306$\cite{bou}}& {\raggedright $1.41^{+0.49}_{-0.10}\cite{bou}$} 
&{\raggedright $6411 \pm 179 \ K\cite{bou}$}
\\
\hline
\end{tabular}
\par}
\centering
\caption
{The stellar radius ($R_{star}$), temperature ($T_{eff}^{star}$) and the 
mean orbital radius ($D$) for the four known
transiting systems.}
\vspace{3mm}
\end{table}

Using, Eq.(\ref{2}) and the parameters from
the above table, we can estimate the 
effective temperatures for the four transiting planets
(we assume an albedo of $a=0.3$\footnote{
Note the dependence of $R_p$ on the albedo 
suggested by Eq.(\ref{1}) is very weak [$\propto (1-a)^{1/8}$]
and for this reason possible uncertainty in albedo does not
significantly affect the prediction for $R_p$.
}).
This information, together with the measured mass
and radius we give in table 2 below:
\begin{table}[h]
\vspace{3mm}
{\centering
\begin{tabular}[t]{|c|c|c|c|}
\hline
{\textbf{Transiting planet}} &
{$R_p$ $[R_J]$} & $M_p$ $[M_J]$ & $T_s$\\
\hline
\hline
{HD209458b} &
{\raggedright $1.43 \pm 0.05$\cite{deeg}}& {\raggedright $0.69 \pm 0.05$
\cite{maz}}
& 1370\ K \\
\hline
{OGLE-TR-56b} &
{\raggedright $1.23 \pm 0.16$}\cite{kon}& {\raggedright $1.45 \pm
0.23$\cite{kon}}
& 1820\ K \\
\hline
{OGLE-TR-113b} &
{\raggedright $1.08 \pm 0.07$\cite{bou}}& {\raggedright $1.35 \pm 0.22$\cite{bou}}
& 1210\ K \\
\hline
{OGLE-TR-132b} &
{\raggedright $1.15^{+0.80}_{-0.13}$\cite{bou}}& {\raggedright $1.01 \pm 0.31$\cite{bou}}
& 1920\ K \\
\hline
\end{tabular}
\par}
\centering
\caption
{The planet radius ($R_p$), mass ($M_p$) and
effective surface temperature ($T_s$)
for the four known transiting planets. For the planet OGLE-TR-113b,
we take the values given in Ref.\cite{bou} (similar results were
independently obtained in Ref.\cite{kon4}).
}
\vspace{3mm}
\end{table}

In figure 1 we plot the values of $R_p$ versus $\sqrt{T_s/M_p}$ for
the three most accurately measured planets, HD209458b, OGLE-TR-56b
and OGLE-TR-113b. The solid line is the prediction, Eq.(\ref{1}),
where we have used HD209458b to fix the proportionality constant.
Clearly, the observations agree remarkably well with
the rough prediction, Eq.(\ref{1}). This appears to be
non-trivial: in the case of ordinary matter
planets, increasing the mass does not significantly affect
the radius, and does not generally lead to a decreasing 
radius (for example, Jupiter is three times heavier than Saturn,
but is 15\% {\it larger}). 
\vskip 0.3cm
\setlength{\unitlength}{0.240900pt}
\ifx\plotpoint\undefined\newsavebox{\plotpoint}\fi
\sbox{\plotpoint}{\rule[-0.200pt]{0.400pt}{0.400pt}}%
\begin{picture}(1500,900)(0,0)
\font\gnuplot=cmr10 at 10pt
\gnuplot
\sbox{\plotpoint}{\rule[-0.200pt]{0.400pt}{0.400pt}}%
\put(181.0,123.0){\rule[-0.200pt]{4.818pt}{0.400pt}}
\put(161,123){\makebox(0,0)[r]{ 0}}
\put(1419.0,123.0){\rule[-0.200pt]{4.818pt}{0.400pt}}
\put(181.0,215.0){\rule[-0.200pt]{4.818pt}{0.400pt}}
\put(161,215){\makebox(0,0)[r]{ 0.2}}
\put(1419.0,215.0){\rule[-0.200pt]{4.818pt}{0.400pt}}
\put(181.0,307.0){\rule[-0.200pt]{4.818pt}{0.400pt}}
\put(161,307){\makebox(0,0)[r]{ 0.4}}
\put(1419.0,307.0){\rule[-0.200pt]{4.818pt}{0.400pt}}
\put(181.0,399.0){\rule[-0.200pt]{4.818pt}{0.400pt}}
\put(161,399){\makebox(0,0)[r]{ 0.6}}
\put(1419.0,399.0){\rule[-0.200pt]{4.818pt}{0.400pt}}
\put(181.0,492.0){\rule[-0.200pt]{4.818pt}{0.400pt}}
\put(161,492){\makebox(0,0)[r]{ 0.8}}
\put(1419.0,492.0){\rule[-0.200pt]{4.818pt}{0.400pt}}
\put(181.0,584.0){\rule[-0.200pt]{4.818pt}{0.400pt}}
\put(161,584){\makebox(0,0)[r]{ 1}}
\put(1419.0,584.0){\rule[-0.200pt]{4.818pt}{0.400pt}}
\put(181.0,676.0){\rule[-0.200pt]{4.818pt}{0.400pt}}
\put(161,676){\makebox(0,0)[r]{ 1.2}}
\put(1419.0,676.0){\rule[-0.200pt]{4.818pt}{0.400pt}}
\put(181.0,768.0){\rule[-0.200pt]{4.818pt}{0.400pt}}
\put(161,768){\makebox(0,0)[r]{ 1.4}}
\put(1419.0,768.0){\rule[-0.200pt]{4.818pt}{0.400pt}}
\put(181.0,860.0){\rule[-0.200pt]{4.818pt}{0.400pt}}
\put(161,860){\makebox(0,0)[r]{ 1.6}}
\put(1419.0,860.0){\rule[-0.200pt]{4.818pt}{0.400pt}}
\put(181.0,123.0){\rule[-0.200pt]{0.400pt}{4.818pt}}
\put(181,82){\makebox(0,0){ 0}}
\put(181.0,840.0){\rule[-0.200pt]{0.400pt}{4.818pt}}
\put(410.0,123.0){\rule[-0.200pt]{0.400pt}{4.818pt}}
\put(410,82){\makebox(0,0){ 0.2}}
\put(410.0,840.0){\rule[-0.200pt]{0.400pt}{4.818pt}}
\put(638.0,123.0){\rule[-0.200pt]{0.400pt}{4.818pt}}
\put(638,82){\makebox(0,0){ 0.4}}
\put(638.0,840.0){\rule[-0.200pt]{0.400pt}{4.818pt}}
\put(867.0,123.0){\rule[-0.200pt]{0.400pt}{4.818pt}}
\put(867,82){\makebox(0,0){ 0.6}}
\put(867.0,840.0){\rule[-0.200pt]{0.400pt}{4.818pt}}
\put(1096.0,123.0){\rule[-0.200pt]{0.400pt}{4.818pt}}
\put(1096,82){\makebox(0,0){ 0.8}}
\put(1096.0,840.0){\rule[-0.200pt]{0.400pt}{4.818pt}}
\put(1325.0,123.0){\rule[-0.200pt]{0.400pt}{4.818pt}}
\put(1325,82){\makebox(0,0){ 1}}
\put(1325.0,840.0){\rule[-0.200pt]{0.400pt}{4.818pt}}
\put(181.0,123.0){\rule[-0.200pt]{303.052pt}{0.400pt}}
\put(1439.0,123.0){\rule[-0.200pt]{0.400pt}{177.543pt}}
\put(181.0,860.0){\rule[-0.200pt]{303.052pt}{0.400pt}}
\put(40,491){\makebox(0,0){$R_p [R_J]$}}
\put(810,21){\makebox(0,0){$\sqrt{T_s/M_p}$}}
\put(181.0,123.0){\rule[-0.200pt]{0.400pt}{177.543pt}}
\put(1325.0,759.0){\rule[-0.200pt]{0.400pt}{11.081pt}}
\put(1315.0,759.0){\rule[-0.200pt]{4.818pt}{0.400pt}}
\put(1315.0,805.0){\rule[-0.200pt]{4.818pt}{0.400pt}}
\put(1090.0,616.0){\rule[-0.200pt]{0.400pt}{35.412pt}}
\put(1080.0,616.0){\rule[-0.200pt]{4.818pt}{0.400pt}}
\put(1080.0,763.0){\rule[-0.200pt]{4.818pt}{0.400pt}}
\put(950.0,588.0){\rule[-0.200pt]{0.400pt}{15.658pt}}
\put(940.0,588.0){\rule[-0.200pt]{4.818pt}{0.400pt}}
\put(940.0,653.0){\rule[-0.200pt]{4.818pt}{0.400pt}}
\put(1285.0,782.0){\rule[-0.200pt]{19.272pt}{0.400pt}}
\put(1285.0,772.0){\rule[-0.200pt]{0.400pt}{4.818pt}}
\put(1365.0,772.0){\rule[-0.200pt]{0.400pt}{4.818pt}}
\put(1033.0,690.0){\rule[-0.200pt]{27.463pt}{0.400pt}}
\put(1033.0,680.0){\rule[-0.200pt]{0.400pt}{4.818pt}}
\put(1147.0,680.0){\rule[-0.200pt]{0.400pt}{4.818pt}}
\put(866.0,620.0){\rule[-0.200pt]{40.230pt}{0.400pt}}
\put(866.0,610.0){\rule[-0.200pt]{0.400pt}{4.818pt}}
\put(1325,782){\raisebox{-.8pt}{\makebox(0,0){$\Diamond$}}}
\put(1090,690){\raisebox{-.8pt}{\makebox(0,0){$\Diamond$}}}
\put(950,620){\raisebox{-.8pt}{\makebox(0,0){$\Diamond$}}}
\put(1033.0,610.0){\rule[-0.200pt]{0.400pt}{4.818pt}}
\put(181.0,123.0){\rule[-0.200pt]{4.818pt}{0.400pt}}
\put(161,123){\makebox(0,0)[r]{ 0}}
\put(1419.0,123.0){\rule[-0.200pt]{4.818pt}{0.400pt}}
\put(181.0,215.0){\rule[-0.200pt]{4.818pt}{0.400pt}}
\put(161,215){\makebox(0,0)[r]{ 0.2}}
\put(1419.0,215.0){\rule[-0.200pt]{4.818pt}{0.400pt}}
\put(181.0,307.0){\rule[-0.200pt]{4.818pt}{0.400pt}}
\put(161,307){\makebox(0,0)[r]{ 0.4}}
\put(1419.0,307.0){\rule[-0.200pt]{4.818pt}{0.400pt}}
\put(181.0,399.0){\rule[-0.200pt]{4.818pt}{0.400pt}}
\put(161,399){\makebox(0,0)[r]{ 0.6}}
\put(1419.0,399.0){\rule[-0.200pt]{4.818pt}{0.400pt}}
\put(181.0,492.0){\rule[-0.200pt]{4.818pt}{0.400pt}}
\put(161,492){\makebox(0,0)[r]{ 0.8}}
\put(1419.0,492.0){\rule[-0.200pt]{4.818pt}{0.400pt}}
\put(181.0,584.0){\rule[-0.200pt]{4.818pt}{0.400pt}}
\put(161,584){\makebox(0,0)[r]{ 1}}
\put(1419.0,584.0){\rule[-0.200pt]{4.818pt}{0.400pt}}
\put(181.0,676.0){\rule[-0.200pt]{4.818pt}{0.400pt}}
\put(161,676){\makebox(0,0)[r]{ 1.2}}
\put(1419.0,676.0){\rule[-0.200pt]{4.818pt}{0.400pt}}
\put(181.0,768.0){\rule[-0.200pt]{4.818pt}{0.400pt}}
\put(161,768){\makebox(0,0)[r]{ 1.4}}
\put(1419.0,768.0){\rule[-0.200pt]{4.818pt}{0.400pt}}
\put(181.0,860.0){\rule[-0.200pt]{4.818pt}{0.400pt}}
\put(161,860){\makebox(0,0)[r]{ 1.6}}
\put(1419.0,860.0){\rule[-0.200pt]{4.818pt}{0.400pt}}
\put(181.0,123.0){\rule[-0.200pt]{0.400pt}{4.818pt}}
\put(181,82){\makebox(0,0){ 0}}
\put(181.0,840.0){\rule[-0.200pt]{0.400pt}{4.818pt}}
\put(410.0,123.0){\rule[-0.200pt]{0.400pt}{4.818pt}}
\put(410,82){\makebox(0,0){ 0.2}}
\put(410.0,840.0){\rule[-0.200pt]{0.400pt}{4.818pt}}
\put(638.0,123.0){\rule[-0.200pt]{0.400pt}{4.818pt}}
\put(638,82){\makebox(0,0){ 0.4}}
\put(638.0,840.0){\rule[-0.200pt]{0.400pt}{4.818pt}}
\put(867.0,123.0){\rule[-0.200pt]{0.400pt}{4.818pt}}
\put(867,82){\makebox(0,0){ 0.6}}
\put(867.0,840.0){\rule[-0.200pt]{0.400pt}{4.818pt}}
\put(1096.0,123.0){\rule[-0.200pt]{0.400pt}{4.818pt}}
\put(1096,82){\makebox(0,0){ 0.8}}
\put(1096.0,840.0){\rule[-0.200pt]{0.400pt}{4.818pt}}
\put(1325.0,123.0){\rule[-0.200pt]{0.400pt}{4.818pt}}
\put(1325,82){\makebox(0,0){ 1}}
\put(1325.0,840.0){\rule[-0.200pt]{0.400pt}{4.818pt}}
\put(181.0,123.0){\rule[-0.200pt]{303.052pt}{0.400pt}}
\put(1439.0,123.0){\rule[-0.200pt]{0.400pt}{177.543pt}}
\put(181.0,860.0){\rule[-0.200pt]{303.052pt}{0.400pt}}
\put(40,491){\makebox(0,0){$R_p [R_J]$}}
\put(810,21){\makebox(0,0){$\sqrt{T_s/M_p}$}}
\put(181.0,123.0){\rule[-0.200pt]{0.400pt}{177.543pt}}
\put(181,123){\usebox{\plotpoint}}
\multiput(181.00,123.59)(0.950,0.485){11}{\rule{0.843pt}{0.117pt}}
\multiput(181.00,122.17)(11.251,7.000){2}{\rule{0.421pt}{0.400pt}}
\multiput(194.00,130.59)(0.758,0.488){13}{\rule{0.700pt}{0.117pt}}
\multiput(194.00,129.17)(10.547,8.000){2}{\rule{0.350pt}{0.400pt}}
\multiput(206.00,138.59)(0.950,0.485){11}{\rule{0.843pt}{0.117pt}}
\multiput(206.00,137.17)(11.251,7.000){2}{\rule{0.421pt}{0.400pt}}
\multiput(219.00,145.59)(0.950,0.485){11}{\rule{0.843pt}{0.117pt}}
\multiput(219.00,144.17)(11.251,7.000){2}{\rule{0.421pt}{0.400pt}}
\multiput(232.00,152.59)(0.824,0.488){13}{\rule{0.750pt}{0.117pt}}
\multiput(232.00,151.17)(11.443,8.000){2}{\rule{0.375pt}{0.400pt}}
\multiput(245.00,160.59)(0.874,0.485){11}{\rule{0.786pt}{0.117pt}}
\multiput(245.00,159.17)(10.369,7.000){2}{\rule{0.393pt}{0.400pt}}
\multiput(257.00,167.59)(0.950,0.485){11}{\rule{0.843pt}{0.117pt}}
\multiput(257.00,166.17)(11.251,7.000){2}{\rule{0.421pt}{0.400pt}}
\multiput(270.00,174.59)(0.824,0.488){13}{\rule{0.750pt}{0.117pt}}
\multiput(270.00,173.17)(11.443,8.000){2}{\rule{0.375pt}{0.400pt}}
\multiput(283.00,182.59)(0.874,0.485){11}{\rule{0.786pt}{0.117pt}}
\multiput(283.00,181.17)(10.369,7.000){2}{\rule{0.393pt}{0.400pt}}
\multiput(295.00,189.59)(0.950,0.485){11}{\rule{0.843pt}{0.117pt}}
\multiput(295.00,188.17)(11.251,7.000){2}{\rule{0.421pt}{0.400pt}}
\multiput(308.00,196.59)(0.824,0.488){13}{\rule{0.750pt}{0.117pt}}
\multiput(308.00,195.17)(11.443,8.000){2}{\rule{0.375pt}{0.400pt}}
\multiput(321.00,204.59)(0.874,0.485){11}{\rule{0.786pt}{0.117pt}}
\multiput(321.00,203.17)(10.369,7.000){2}{\rule{0.393pt}{0.400pt}}
\multiput(333.00,211.59)(0.950,0.485){11}{\rule{0.843pt}{0.117pt}}
\multiput(333.00,210.17)(11.251,7.000){2}{\rule{0.421pt}{0.400pt}}
\multiput(346.00,218.59)(0.950,0.485){11}{\rule{0.843pt}{0.117pt}}
\multiput(346.00,217.17)(11.251,7.000){2}{\rule{0.421pt}{0.400pt}}
\multiput(359.00,225.59)(0.824,0.488){13}{\rule{0.750pt}{0.117pt}}
\multiput(359.00,224.17)(11.443,8.000){2}{\rule{0.375pt}{0.400pt}}
\multiput(372.00,233.59)(0.874,0.485){11}{\rule{0.786pt}{0.117pt}}
\multiput(372.00,232.17)(10.369,7.000){2}{\rule{0.393pt}{0.400pt}}
\multiput(384.00,240.59)(0.950,0.485){11}{\rule{0.843pt}{0.117pt}}
\multiput(384.00,239.17)(11.251,7.000){2}{\rule{0.421pt}{0.400pt}}
\multiput(397.00,247.59)(0.824,0.488){13}{\rule{0.750pt}{0.117pt}}
\multiput(397.00,246.17)(11.443,8.000){2}{\rule{0.375pt}{0.400pt}}
\multiput(410.00,255.59)(0.874,0.485){11}{\rule{0.786pt}{0.117pt}}
\multiput(410.00,254.17)(10.369,7.000){2}{\rule{0.393pt}{0.400pt}}
\multiput(422.00,262.59)(0.950,0.485){11}{\rule{0.843pt}{0.117pt}}
\multiput(422.00,261.17)(11.251,7.000){2}{\rule{0.421pt}{0.400pt}}
\multiput(435.00,269.59)(0.824,0.488){13}{\rule{0.750pt}{0.117pt}}
\multiput(435.00,268.17)(11.443,8.000){2}{\rule{0.375pt}{0.400pt}}
\multiput(448.00,277.59)(0.950,0.485){11}{\rule{0.843pt}{0.117pt}}
\multiput(448.00,276.17)(11.251,7.000){2}{\rule{0.421pt}{0.400pt}}
\multiput(461.00,284.59)(0.874,0.485){11}{\rule{0.786pt}{0.117pt}}
\multiput(461.00,283.17)(10.369,7.000){2}{\rule{0.393pt}{0.400pt}}
\multiput(473.00,291.59)(0.824,0.488){13}{\rule{0.750pt}{0.117pt}}
\multiput(473.00,290.17)(11.443,8.000){2}{\rule{0.375pt}{0.400pt}}
\multiput(486.00,299.59)(0.950,0.485){11}{\rule{0.843pt}{0.117pt}}
\multiput(486.00,298.17)(11.251,7.000){2}{\rule{0.421pt}{0.400pt}}
\multiput(499.00,306.59)(0.874,0.485){11}{\rule{0.786pt}{0.117pt}}
\multiput(499.00,305.17)(10.369,7.000){2}{\rule{0.393pt}{0.400pt}}
\multiput(511.00,313.59)(0.824,0.488){13}{\rule{0.750pt}{0.117pt}}
\multiput(511.00,312.17)(11.443,8.000){2}{\rule{0.375pt}{0.400pt}}
\multiput(524.00,321.59)(0.950,0.485){11}{\rule{0.843pt}{0.117pt}}
\multiput(524.00,320.17)(11.251,7.000){2}{\rule{0.421pt}{0.400pt}}
\multiput(537.00,328.59)(0.950,0.485){11}{\rule{0.843pt}{0.117pt}}
\multiput(537.00,327.17)(11.251,7.000){2}{\rule{0.421pt}{0.400pt}}
\multiput(550.00,335.59)(0.758,0.488){13}{\rule{0.700pt}{0.117pt}}
\multiput(550.00,334.17)(10.547,8.000){2}{\rule{0.350pt}{0.400pt}}
\multiput(562.00,343.59)(0.950,0.485){11}{\rule{0.843pt}{0.117pt}}
\multiput(562.00,342.17)(11.251,7.000){2}{\rule{0.421pt}{0.400pt}}
\multiput(575.00,350.59)(0.950,0.485){11}{\rule{0.843pt}{0.117pt}}
\multiput(575.00,349.17)(11.251,7.000){2}{\rule{0.421pt}{0.400pt}}
\multiput(588.00,357.59)(0.758,0.488){13}{\rule{0.700pt}{0.117pt}}
\multiput(588.00,356.17)(10.547,8.000){2}{\rule{0.350pt}{0.400pt}}
\multiput(600.00,365.59)(0.950,0.485){11}{\rule{0.843pt}{0.117pt}}
\multiput(600.00,364.17)(11.251,7.000){2}{\rule{0.421pt}{0.400pt}}
\multiput(613.00,372.59)(0.950,0.485){11}{\rule{0.843pt}{0.117pt}}
\multiput(613.00,371.17)(11.251,7.000){2}{\rule{0.421pt}{0.400pt}}
\multiput(626.00,379.59)(0.874,0.485){11}{\rule{0.786pt}{0.117pt}}
\multiput(626.00,378.17)(10.369,7.000){2}{\rule{0.393pt}{0.400pt}}
\multiput(638.00,386.59)(0.824,0.488){13}{\rule{0.750pt}{0.117pt}}
\multiput(638.00,385.17)(11.443,8.000){2}{\rule{0.375pt}{0.400pt}}
\multiput(651.00,394.59)(0.950,0.485){11}{\rule{0.843pt}{0.117pt}}
\multiput(651.00,393.17)(11.251,7.000){2}{\rule{0.421pt}{0.400pt}}
\multiput(664.00,401.59)(0.950,0.485){11}{\rule{0.843pt}{0.117pt}}
\multiput(664.00,400.17)(11.251,7.000){2}{\rule{0.421pt}{0.400pt}}
\multiput(677.00,408.59)(0.758,0.488){13}{\rule{0.700pt}{0.117pt}}
\multiput(677.00,407.17)(10.547,8.000){2}{\rule{0.350pt}{0.400pt}}
\multiput(689.00,416.59)(0.950,0.485){11}{\rule{0.843pt}{0.117pt}}
\multiput(689.00,415.17)(11.251,7.000){2}{\rule{0.421pt}{0.400pt}}
\multiput(702.00,423.59)(0.950,0.485){11}{\rule{0.843pt}{0.117pt}}
\multiput(702.00,422.17)(11.251,7.000){2}{\rule{0.421pt}{0.400pt}}
\multiput(715.00,430.59)(0.758,0.488){13}{\rule{0.700pt}{0.117pt}}
\multiput(715.00,429.17)(10.547,8.000){2}{\rule{0.350pt}{0.400pt}}
\multiput(727.00,438.59)(0.950,0.485){11}{\rule{0.843pt}{0.117pt}}
\multiput(727.00,437.17)(11.251,7.000){2}{\rule{0.421pt}{0.400pt}}
\multiput(740.00,445.59)(0.950,0.485){11}{\rule{0.843pt}{0.117pt}}
\multiput(740.00,444.17)(11.251,7.000){2}{\rule{0.421pt}{0.400pt}}
\multiput(753.00,452.59)(0.824,0.488){13}{\rule{0.750pt}{0.117pt}}
\multiput(753.00,451.17)(11.443,8.000){2}{\rule{0.375pt}{0.400pt}}
\multiput(766.00,460.59)(0.874,0.485){11}{\rule{0.786pt}{0.117pt}}
\multiput(766.00,459.17)(10.369,7.000){2}{\rule{0.393pt}{0.400pt}}
\multiput(778.00,467.59)(0.950,0.485){11}{\rule{0.843pt}{0.117pt}}
\multiput(778.00,466.17)(11.251,7.000){2}{\rule{0.421pt}{0.400pt}}
\multiput(791.00,474.59)(0.824,0.488){13}{\rule{0.750pt}{0.117pt}}
\multiput(791.00,473.17)(11.443,8.000){2}{\rule{0.375pt}{0.400pt}}
\multiput(804.00,482.59)(0.874,0.485){11}{\rule{0.786pt}{0.117pt}}
\multiput(804.00,481.17)(10.369,7.000){2}{\rule{0.393pt}{0.400pt}}
\multiput(816.00,489.59)(0.950,0.485){11}{\rule{0.843pt}{0.117pt}}
\multiput(816.00,488.17)(11.251,7.000){2}{\rule{0.421pt}{0.400pt}}
\multiput(829.00,496.59)(0.824,0.488){13}{\rule{0.750pt}{0.117pt}}
\multiput(829.00,495.17)(11.443,8.000){2}{\rule{0.375pt}{0.400pt}}
\multiput(842.00,504.59)(0.874,0.485){11}{\rule{0.786pt}{0.117pt}}
\multiput(842.00,503.17)(10.369,7.000){2}{\rule{0.393pt}{0.400pt}}
\multiput(854.00,511.59)(0.950,0.485){11}{\rule{0.843pt}{0.117pt}}
\multiput(854.00,510.17)(11.251,7.000){2}{\rule{0.421pt}{0.400pt}}
\multiput(867.00,518.59)(0.824,0.488){13}{\rule{0.750pt}{0.117pt}}
\multiput(867.00,517.17)(11.443,8.000){2}{\rule{0.375pt}{0.400pt}}
\multiput(880.00,526.59)(0.950,0.485){11}{\rule{0.843pt}{0.117pt}}
\multiput(880.00,525.17)(11.251,7.000){2}{\rule{0.421pt}{0.400pt}}
\multiput(893.00,533.59)(0.874,0.485){11}{\rule{0.786pt}{0.117pt}}
\multiput(893.00,532.17)(10.369,7.000){2}{\rule{0.393pt}{0.400pt}}
\multiput(905.00,540.59)(0.950,0.485){11}{\rule{0.843pt}{0.117pt}}
\multiput(905.00,539.17)(11.251,7.000){2}{\rule{0.421pt}{0.400pt}}
\multiput(918.00,547.59)(0.824,0.488){13}{\rule{0.750pt}{0.117pt}}
\multiput(918.00,546.17)(11.443,8.000){2}{\rule{0.375pt}{0.400pt}}
\multiput(931.00,555.59)(0.874,0.485){11}{\rule{0.786pt}{0.117pt}}
\multiput(931.00,554.17)(10.369,7.000){2}{\rule{0.393pt}{0.400pt}}
\multiput(943.00,562.59)(0.950,0.485){11}{\rule{0.843pt}{0.117pt}}
\multiput(943.00,561.17)(11.251,7.000){2}{\rule{0.421pt}{0.400pt}}
\multiput(956.00,569.59)(0.824,0.488){13}{\rule{0.750pt}{0.117pt}}
\multiput(956.00,568.17)(11.443,8.000){2}{\rule{0.375pt}{0.400pt}}
\multiput(969.00,577.59)(0.950,0.485){11}{\rule{0.843pt}{0.117pt}}
\multiput(969.00,576.17)(11.251,7.000){2}{\rule{0.421pt}{0.400pt}}
\multiput(982.00,584.59)(0.874,0.485){11}{\rule{0.786pt}{0.117pt}}
\multiput(982.00,583.17)(10.369,7.000){2}{\rule{0.393pt}{0.400pt}}
\multiput(994.00,591.59)(0.824,0.488){13}{\rule{0.750pt}{0.117pt}}
\multiput(994.00,590.17)(11.443,8.000){2}{\rule{0.375pt}{0.400pt}}
\multiput(1007.00,599.59)(0.950,0.485){11}{\rule{0.843pt}{0.117pt}}
\multiput(1007.00,598.17)(11.251,7.000){2}{\rule{0.421pt}{0.400pt}}
\multiput(1020.00,606.59)(0.874,0.485){11}{\rule{0.786pt}{0.117pt}}
\multiput(1020.00,605.17)(10.369,7.000){2}{\rule{0.393pt}{0.400pt}}
\multiput(1032.00,613.59)(0.824,0.488){13}{\rule{0.750pt}{0.117pt}}
\multiput(1032.00,612.17)(11.443,8.000){2}{\rule{0.375pt}{0.400pt}}
\multiput(1045.00,621.59)(0.950,0.485){11}{\rule{0.843pt}{0.117pt}}
\multiput(1045.00,620.17)(11.251,7.000){2}{\rule{0.421pt}{0.400pt}}
\multiput(1058.00,628.59)(0.874,0.485){11}{\rule{0.786pt}{0.117pt}}
\multiput(1058.00,627.17)(10.369,7.000){2}{\rule{0.393pt}{0.400pt}}
\multiput(1070.00,635.59)(0.824,0.488){13}{\rule{0.750pt}{0.117pt}}
\multiput(1070.00,634.17)(11.443,8.000){2}{\rule{0.375pt}{0.400pt}}
\multiput(1083.00,643.59)(0.950,0.485){11}{\rule{0.843pt}{0.117pt}}
\multiput(1083.00,642.17)(11.251,7.000){2}{\rule{0.421pt}{0.400pt}}
\multiput(1096.00,650.59)(0.950,0.485){11}{\rule{0.843pt}{0.117pt}}
\multiput(1096.00,649.17)(11.251,7.000){2}{\rule{0.421pt}{0.400pt}}
\multiput(1109.00,657.59)(0.758,0.488){13}{\rule{0.700pt}{0.117pt}}
\multiput(1109.00,656.17)(10.547,8.000){2}{\rule{0.350pt}{0.400pt}}
\multiput(1121.00,665.59)(0.950,0.485){11}{\rule{0.843pt}{0.117pt}}
\multiput(1121.00,664.17)(11.251,7.000){2}{\rule{0.421pt}{0.400pt}}
\multiput(1134.00,672.59)(0.950,0.485){11}{\rule{0.843pt}{0.117pt}}
\multiput(1134.00,671.17)(11.251,7.000){2}{\rule{0.421pt}{0.400pt}}
\multiput(1147.00,679.59)(0.758,0.488){13}{\rule{0.700pt}{0.117pt}}
\multiput(1147.00,678.17)(10.547,8.000){2}{\rule{0.350pt}{0.400pt}}
\multiput(1159.00,687.59)(0.950,0.485){11}{\rule{0.843pt}{0.117pt}}
\multiput(1159.00,686.17)(11.251,7.000){2}{\rule{0.421pt}{0.400pt}}
\multiput(1172.00,694.59)(0.950,0.485){11}{\rule{0.843pt}{0.117pt}}
\multiput(1172.00,693.17)(11.251,7.000){2}{\rule{0.421pt}{0.400pt}}
\multiput(1185.00,701.59)(0.824,0.488){13}{\rule{0.750pt}{0.117pt}}
\multiput(1185.00,700.17)(11.443,8.000){2}{\rule{0.375pt}{0.400pt}}
\multiput(1198.00,709.59)(0.874,0.485){11}{\rule{0.786pt}{0.117pt}}
\multiput(1198.00,708.17)(10.369,7.000){2}{\rule{0.393pt}{0.400pt}}
\multiput(1210.00,716.59)(0.950,0.485){11}{\rule{0.843pt}{0.117pt}}
\multiput(1210.00,715.17)(11.251,7.000){2}{\rule{0.421pt}{0.400pt}}
\multiput(1223.00,723.59)(0.950,0.485){11}{\rule{0.843pt}{0.117pt}}
\multiput(1223.00,722.17)(11.251,7.000){2}{\rule{0.421pt}{0.400pt}}
\multiput(1236.00,730.59)(0.758,0.488){13}{\rule{0.700pt}{0.117pt}}
\multiput(1236.00,729.17)(10.547,8.000){2}{\rule{0.350pt}{0.400pt}}
\multiput(1248.00,738.59)(0.950,0.485){11}{\rule{0.843pt}{0.117pt}}
\multiput(1248.00,737.17)(11.251,7.000){2}{\rule{0.421pt}{0.400pt}}
\multiput(1261.00,745.59)(0.950,0.485){11}{\rule{0.843pt}{0.117pt}}
\multiput(1261.00,744.17)(11.251,7.000){2}{\rule{0.421pt}{0.400pt}}
\multiput(1274.00,752.59)(0.824,0.488){13}{\rule{0.750pt}{0.117pt}}
\multiput(1274.00,751.17)(11.443,8.000){2}{\rule{0.375pt}{0.400pt}}
\multiput(1287.00,760.59)(0.874,0.485){11}{\rule{0.786pt}{0.117pt}}
\multiput(1287.00,759.17)(10.369,7.000){2}{\rule{0.393pt}{0.400pt}}
\multiput(1299.00,767.59)(0.950,0.485){11}{\rule{0.843pt}{0.117pt}}
\multiput(1299.00,766.17)(11.251,7.000){2}{\rule{0.421pt}{0.400pt}}
\multiput(1312.00,774.59)(0.824,0.488){13}{\rule{0.750pt}{0.117pt}}
\multiput(1312.00,773.17)(11.443,8.000){2}{\rule{0.375pt}{0.400pt}}
\multiput(1325.00,782.59)(0.874,0.485){11}{\rule{0.786pt}{0.117pt}}
\multiput(1325.00,781.17)(10.369,7.000){2}{\rule{0.393pt}{0.400pt}}
\multiput(1337.00,789.59)(0.950,0.485){11}{\rule{0.843pt}{0.117pt}}
\multiput(1337.00,788.17)(11.251,7.000){2}{\rule{0.421pt}{0.400pt}}
\multiput(1350.00,796.59)(0.824,0.488){13}{\rule{0.750pt}{0.117pt}}
\multiput(1350.00,795.17)(11.443,8.000){2}{\rule{0.375pt}{0.400pt}}
\multiput(1363.00,804.59)(0.874,0.485){11}{\rule{0.786pt}{0.117pt}}
\multiput(1363.00,803.17)(10.369,7.000){2}{\rule{0.393pt}{0.400pt}}
\multiput(1375.00,811.59)(0.950,0.485){11}{\rule{0.843pt}{0.117pt}}
\multiput(1375.00,810.17)(11.251,7.000){2}{\rule{0.421pt}{0.400pt}}
\multiput(1388.00,818.59)(0.824,0.488){13}{\rule{0.750pt}{0.117pt}}
\multiput(1388.00,817.17)(11.443,8.000){2}{\rule{0.375pt}{0.400pt}}
\multiput(1401.00,826.59)(0.950,0.485){11}{\rule{0.843pt}{0.117pt}}
\multiput(1401.00,825.17)(11.251,7.000){2}{\rule{0.421pt}{0.400pt}}
\multiput(1414.00,833.59)(0.874,0.485){11}{\rule{0.786pt}{0.117pt}}
\multiput(1414.00,832.17)(10.369,7.000){2}{\rule{0.393pt}{0.400pt}}
\multiput(1426.00,840.59)(0.824,0.488){13}{\rule{0.750pt}{0.117pt}}
\multiput(1426.00,839.17)(11.443,8.000){2}{\rule{0.375pt}{0.400pt}}
\end{picture}

\vskip 0.5cm
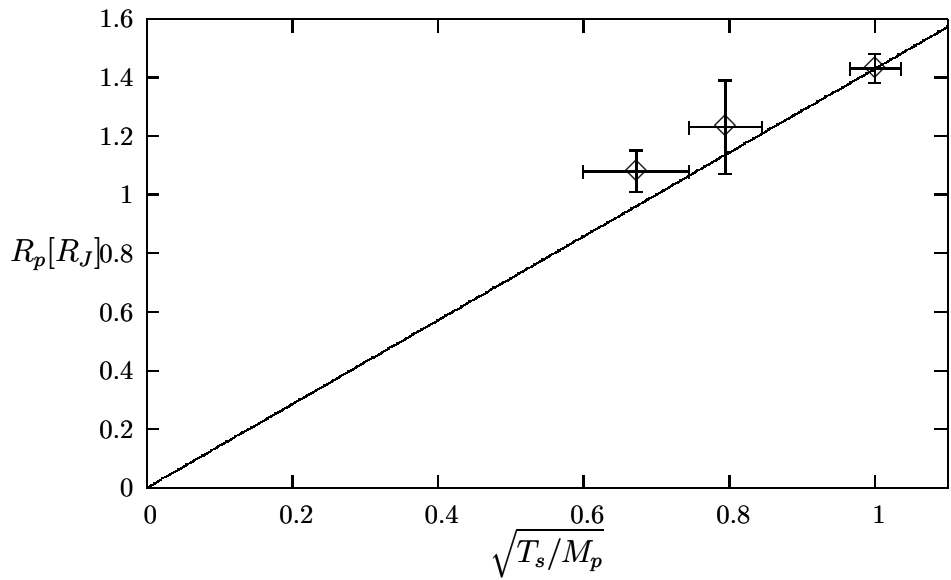
{\small \noindent Figure 1: The measured effective size, $R_p$, of
the transiting planets HD209458b, OGLE-TR-56b and OGLE-TR-113b
versus $\sqrt{T_s/M_p}$ (in units where $\sqrt{T_s/M_p} = 1$
for HD209458b). The straightline
is the prediction, Eq.(\ref{1}), which assumes that
the planets are 
composed predominately of mirror matter.}
\vskip 1.0cm

In conclusion, we have compared the rough prediction, Eq.(\ref{1}),
with the recently discovered close-in transiting 
planets OGLE-TR-56b and OLGE-TR-113b. This prediction is found 
to be in agreement with the observations which seems to 
favour the mirror matter interpretation of the close-in extrasolar
planets.  However, it is certainly possible that the apparent agreement 
with the rough predicition, Eq.(\ref{1}) is coincidental -- so 
more data would be welcome. 
Especially decisive would be
the discovery of a much heavier transiting planet,
$M_p \stackrel{>}{\sim} 2M_J$,
which should have a radius less than $R_J$ if it is
a mirror world.

\end{document}